\documentclass[useAMS,usenatbib]{mn2e}

\usepackage{graphicx}

\usepackage{epstopdf}    %%%%%% CJS ADDED THIS TO ENABLE EPS FIGURES ************

\title[H$\alpha$ emission with AKARI-FUHYU near-infrared spectroscopy]{Detection of H$\alpha$ emission from $z > 3.5$ submillimetre luminous galaxies with {\it AKARI}-FUHYU spectroscopy}

\author[Chris Sedgwick et al.]
{Chris Sedgwick$^{1}$,
 Stephen Serjeant$^{1}$, 
 Chris Pearson$^{2,1}$, 
 Ian Smail$^{3}$, 
 Myungshin Im$^{4}$,
 \newauthor
 Shinki Oyabu$^{6}$, 
 Toshinobu Takagi$^{5}$, 
 Hideo Matsuhara$^{5}$,  
 Takehiko Wada$^{5}$,
 \newauthor
 Hyung Mok Lee$^{4}$, 
 Woong-Seob Jeong$^{7}$ 
 and Glenn J. White$^{1,2}$\\
$^{1}$Department of Physical Sciences, The Open University, Milton Keynes MK7 6AA, UK\\
$^{2}$RAL Space, Rutherford Appleton Laboratory, Chilton, Didcot, Oxfordshire OX11 0QX, UK\\
$^{3}$Institute for Computational Cosmology, Durham University, Durham DH1 3LE, UK\\
$^{4}$Department of Physics and Astronomy, Seoul National University, Seoul 151-742, Korea\\
$^{5}$Institute of Space and Astronautical Science, Japan Aerospace Exploration Agency, Sagamihara, Kanagawa, 229 8510, Japan\\
$^{6}$Graduate School of Science, Nagoya University, Furo-cho, Chikusa-ku, Nagoya, Aichi 464-8602, Japan\\
$^{7}$Korea Astronomy \& Space Science Institute, 776 Daedeok-daero, Yuseong-gu, Daejeon 305-348, Korea}

\begin{document}

\date{Re-submitted June 2013}

\pagerange{\pageref{firstpage}--\pageref{lastpage}} \pubyear{2010}

\maketitle

\label{firstpage}

\begin{abstract}
We present tentative H$\alpha$ emission line detections of four submillimetre-detected galaxies at $z>3.5$: the radio galaxies 8C1909+722 and 4C60.07 at signal-to-noise ratios (SNRs) of 3.1 and 2.5, and two submillimetre-selected galaxies (SMGs) near the first of these at SNRs of 10.0 and 2.4, made with the {\it AKARI Space Telescope} as part of the FUHYU mission program. These are the highest-redshift H$\alpha$ detections in such galaxies, made possible by {\it AKARI's} unique near-infrared spectroscopic capability. The two radio galaxies had known redshifts and surrounding structure, and we have detected broad H$\alpha$ components indicating the presence of dust-shrouded quasars. We conclude that powerful AGNs at $z>3.5$ occur in peaks of the star-formation density fields, supporting a close connection between stellar mass build-up and black hole mass assembly at this redshift. We also show that 4C60.07 is a binary AGN. The H$\alpha$  detections of the two SMGs are the first redshift determinations for these sources, confirming their physical association around their companion radio galaxy. The H$\alpha$-derived star formation rates (SFRs) for the SMGs are lower than their far-infrared derived SFRs by a factor of $\sim$10, suggesting a level of dust obscuration similar to that found in studies at $\sim1<z<2.7$. \end{abstract}


\begin{table*}
 \caption{Sample of  high-redshift radio galaxies and submillimetre-selected sources. The 850 $\mu$m flux data from SCUBA and the target coordinates are from Stevens et al. (2003).} \label{table:targets}
  \begin{center}
   \scalebox{0.9}{
    \begin{tabular}{|lccccccr}
\hline
                                                      &                             &                         &                                         &   850 $\mu$m                                  & \multicolumn{3}{c}{--- AKARI NIR Spectroscopy ---}                     \\
                                                       &    RA                   &    Dec              & Redshift                         &   Flux                           & Observation    &                                & Exposure     \\ 
 Source                                         & (J2000)              &  (J2000)         & (z)                                    &   (mJy)                       &      ID                 & Dates                     & (mins)                                   \\
 \hline
8C1909+722 HzRG                        &  19 08 23.3     & +72 20 10.4    & 3.536 $\pm$0.0003  & 34.9 $\pm$3.0         & 1370153       &    8/2008(1), 8/2009(9)    & ~93   \\
SMMJ190827+721928 (SMM1)   &  19 08 27.4     & +72 19 28.0     &   ...                    & 23.0 $\pm$2.5         & 1370154       &   8/2008(10)                  & ~93\\
SMMJ190829+722050 (SMM2)   &  19 08 29.3     & +72 20 49.6     &   ...                   & ~8.7 $\pm$2.4          & 1370155       &     8/2009(10)                & ~93\\
SMMJ190816+722024 (SMM3)   &  19 08 16.1     & +72 20 24.0     &  ...                   &  ~4.3 $\pm$2.1         & 1370156       &    8/2008(5), 2/2009(2),   & \\
                                                            &                         &                              &                                    &                                     &                        &    8/2009(3), 2/2010(3)     & 120\\ 
4C60.07         HzRG                        &  05 12 54.8     & +60 30 51.7       & 3.788 $\pm$0.004   &  23.8 $\pm$3.5        & 1370162       &     6/2008(9), 6/2009(1)   & ~93     \\
 \hline
   \end{tabular}
}
\end{center}
\end{table*}

\section{Introduction}

The interaction between active galactic nuclei (AGNs) and star formation in galaxies is a key question in cosmology. For local galaxies, Magorrian et al.\ (1998) found a relation between black hole mass and stellar bulge mass, and Ferrarese \& Merritt (2000) and Gebhardt et al.\ (2000) found a tight correlation between black hole mass and stellar velocity dispersion, so it is widely accepted that the two are intimately related. The redshift evolution of the cosmic star formation rate (SFR) is remarkably similar to that of quasar luminosity density (Boyle \& Terlevich 1998; Chapman et al.\ 2005) and black hole accretion (Franceschini et al.\ 1999). Mid-infrared spectroscopy has shown that most Ultra-Luminous InfraRed Galaxies (ULIRGs, $10^{12}L_{\sun}<L_{IR}<10^{13}L_{\sun}$) have simultaneous AGN and starburst activity in their nuclei (Genzel et al.\ 1998; Spoon et al.\ 2007).

Various models have been developed to explain the dynamics underlying these observations. Simulations of galaxy mergers have incorporated the growth of black holes and star formation, showing AGN feedback as a mechanism to regulate SFR (e.g.\ Springel et al.\ 2005; di Matteo et al.\ 2005). Semi-analytic source count models can best reproduce massive galaxy number densities when incorporating AGN feedback in the co-evolution of galaxies and their central black holes (e.g.\ Croton et al.\ 2006; Bower et al.\ 2006). 

Observational constraints for such models from high-redshift galaxies are now becoming available. Studies using a combination of submillimetre and X-ray observations at $z \sim 2$ have confirmed the association of AGNs and intense star formation (Alexander et al.\ 2005; Harrison et al.\ 2012).

The discovery of high-redshift submillimetre galaxies (Smail et al.\ 1997; Hughes et al.\ 1998; Barger et al.\ 1998) and their possible association with high-redshift radio galaxies (HzRGs) (e.g.\ Stevens et al.\ 2003) implies that there are regions of intense star-formation which also show strong radio emission at high redshift. Detection of emission lines from these galaxies in the infrared can be used to confirm this connection, and to provide an independent measure of SFRs and AGN activity.

Spectroscopy of high-redshift radio and submillimetre galaxies has previously been made in the near-infrared K-band (Swinbank et al. 2004). The {\it AKARI Space Telescope} (Murakami et al.\ 2007) could obtain spectra at longer near-infrared wavelengths between 2.5 $\mu$m and 5.0 $\mu$m, a region not previously available for spectroscopy by other infrared space missions such as  {\it Spitzer Space Telescope} (and with poor sensitivity from the ground), giving the possibility of detecting H$\alpha$ for galaxies at  $3.0<z<6.5$. 

This paper presents tentative H$\alpha$ detections from {\it AKARI} observations of high-redshift radio and submillimetre galaxies. Data collection and reduction are discussed in \S \ref{sec:methods}. Results are presented in \S  \ref{sec:results} and discussed in \S \ref{sec:discussion}. We assume $H_0$=72.0 km s$^-$$^1$Mpc$^-$$^1$, $\Omega_M$=0.3 and $\Omega_{\Lambda}$=0.7.

\begin{figure*}
  \begin{center}  
    \resizebox{2.35in}{!}{\includegraphics{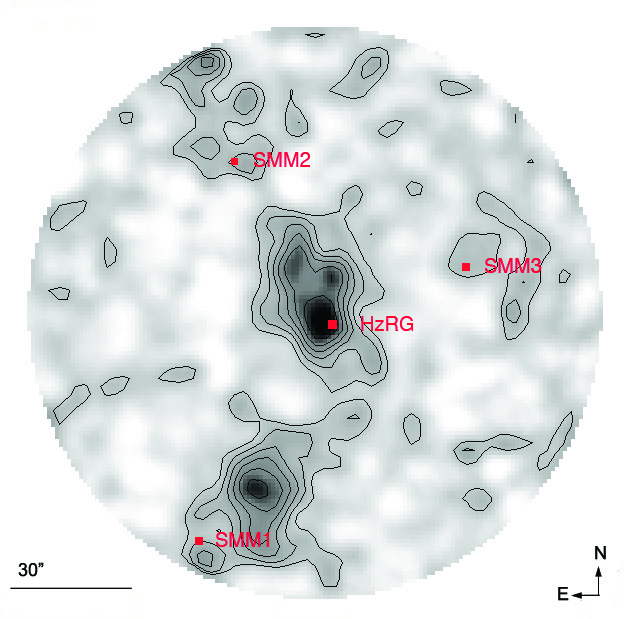}}  
          \resizebox{0.2in}{!}{\includegraphics{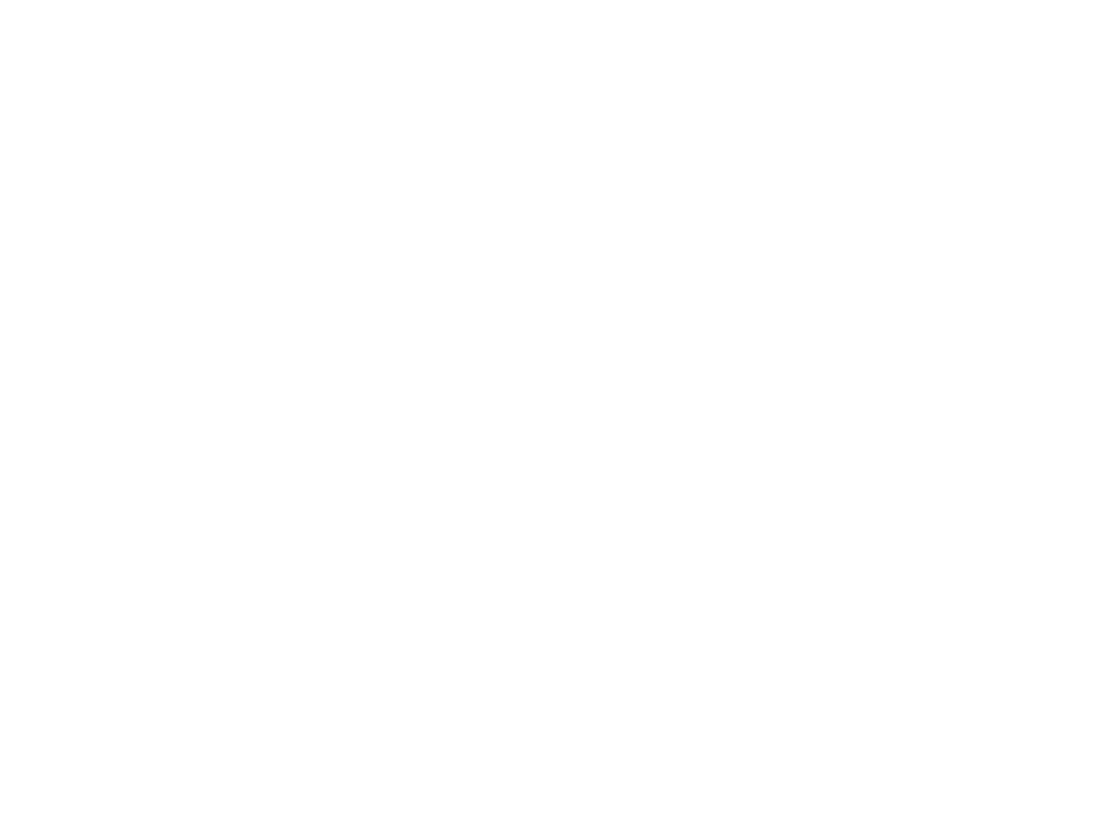}} % white space
        \resizebox{2.65in}{!}{\includegraphics{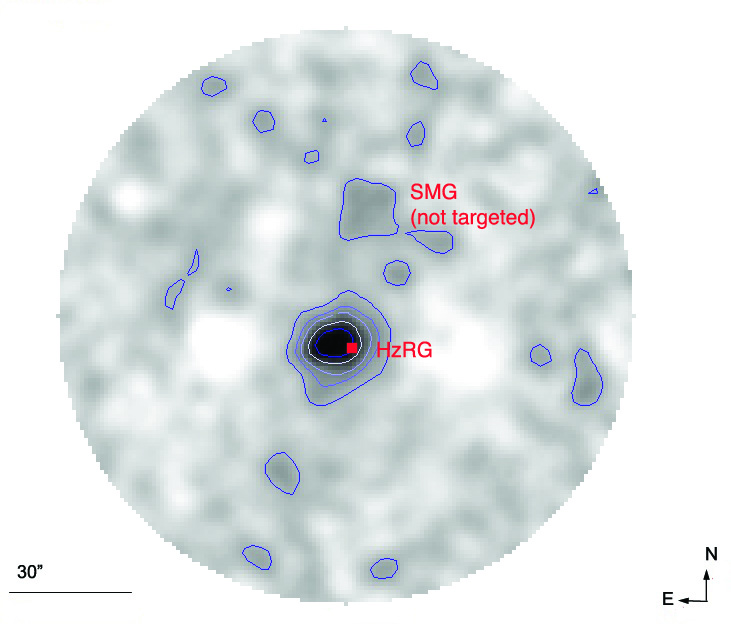}} 
\caption{\small Left: SCUBA 850 $\mu$m image and contours (at 0.5, 1, 1.5, 2, 2.5 and 3 $\sigma$) for 8C1909+722 with the positions of the four {\it AKARI} spectra marked in red. Right: SCUBA 850 $\mu$m image and contours (at 1, 2, 3 and 4 $\sigma$) for 4C 60.07 with the position of the {\it AKARI} spectrum marked in red. This does not resolve the sources within the central contour (see Figure \ref{fig:images_blowup} for SMA 870 $\mu$m contours). The SCUBA maps are from Stevens et al.\ (2003).} \label{fig:images_submm}
 \end{center} 
\end{figure*}

%\vspace{-10 pt}

\section{Data Collection and Reduction}\label{sec:methods}

\subsection{{\it AKARI}-FUHYU Mission Program}

The {\it AKARI}  \textquotedblleft Follow-Up Hayai-Yasui-Umai\textquotedblright~mission program (FUHYU, Pearson et al.\ 2010) used the {\it AKARI} InfraRed Camera (IRC, Onaka et al.\ 2007) to carry out imaging and spectroscopy of well-studied galaxies rich in multi-wavelength data, often with known redshifts, to maximise the legacy value of the {\it AKARI} data. The mission program carried out extensive infrared imaging during the first two phases of the {\it AKARI} mission; the near-infrared spectroscopic campaign was carried out during Phase III, the warm phase after cryogen exhaustion. Spectroscopy was carried out using the IRC grism over a 1$\arcmin~\times $ 1$\arcmin$ point source aperture centred on 46 submillimetre and radio galaxies from June 2008 - May 2010. We will be reporting later on results for the rest of our FUHYU mission program targets.% (Sedgwick et al.\ in preparation).

%A key objective was to detect  potentially powerful H$\alpha$ emission in high redshift sources. 

%, in the Lockman Hole field (Serjeant et al.\ 2009), and in the GOODS-N field (Pearson et al.\ 2010; Negrello et al.\ 2009). 

%A later part of the spectroscopic program observed 26 {\it Spitzer} 70 $\mu$m selected galaxies. We obtained a total of 552 pointings, each with $\sim 9.3$ minutes exposure across these 72 sources.

\vspace{-10 pt}

\subsection{Targets Observed}\label{sec:previous}

Observations were made of two HzRGs, and three
submillimetre sources potentially associated with one of them (see Table \ref{table:targets} and Figure \ref{fig:images_submm}). Stevens et al.\ (2003) described submillimetre mapping of these radio sources by the Submillimeter Common-User Bolometer Array (SCUBA) instrument on the James Clerk Maxwell Telescope (JCMT), and suggested that the submillimetre sources observed were associated with the nearby radio source on the basis of their number densities and positions relative to the radio jets.

 \vspace{4pt}
 
\emph{8C1909+722 and companions.}  
The redshift was reported  in De Breuck et al.\ (2001) based on a deep Keck optical spectrum showing strong 
Ly-$\alpha$ emission with FWHM = 1200 $\pm$ 90 km s$^{-1}$. This field was included in an 850 $\mu$m survey of the environments of HzRGs which detected three nearby submillimetre galaxies (Stevens et al.\ 2003). 
Two of the three submillimetre sources (SMM1 and SMM2) were later detected at 350 $\mu$m with SHARC-II on the Caltech Submillimeter Observatory (Greve et al.\ 2006). A recent study using JVLA and IRAM PdBI radio observations and Herschel data at 100 $\mu$m - 500 $\mu$m (Ivison et al.\ 2012) found a large red dust feature aligned with the radio jet and SMM1, and  concluded that SMM1 probably shared the same node or filament of the cosmic web as the radio galaxy, although it did not detect convincing $^{12}$CO emission from SMM1. Redshifts for these SMGs were not previously known.

\vspace{4pt}

\emph{4C60.07.} The redshift for this galaxy is based on the Ly-$\alpha$ emission line (Roettgering et al.\ 1997) with  FWHM = 2880 $\pm$ 940 km s$^{-1}$. Dust emission was detected at 850 $\mu$m and 1.25 mm (Papadopoulos et al.\ 2000), and CO J=1-0 emission (Greve et al.\ 2004). A detailed SMA, {\it Spitzer} and VLA study by Ivison et al.\ (2008) suggested an early-stage merger between the host galaxy of an AGN (the HzRG) and a companion starburst/AGN (which they labelled \textquoteleft B\textquoteright; see Figure \ref{fig:images_blowup}). They proposed that a second submillimetre source, which they labelled \textquoteleft A\textquoteright , although of roughly equal integrated flux as source  \textquoteleft B\textquoteright , might be comprised of cold dust and gas, a short-lived tidal structure caused by the merger.% A further submillimetre source may be associated with this HzRG (Stevens et al. 2003; see Figure \ref{fig:images_blowup}, middle) but was not targeted in this study.

\subsection{Observations}

The {\it AKARI} IRC-NIR instrument used a filter wheel to select either one of three imaging filters or one of two dispersion elements. For our reference image, we used the {\it N3} image filter ($\sim$ 3 $\mu$m) which had a field of view of approximately $10'\times10'$ across $412\times512$ pixels giving a pixel scale of $1.46''$. For spectroscopy, we used the grism {\it NG} which dispersed between 2.5 $\mu$m - 5.0 $\mu$m across 291 pixels, giving 0.0097 $\mu$m per pixel, in a 1\arcmin $\times$1\arcmin~aperture (referred to as {\it Np}) which is dedicated to spectroscopy of point sources. The PSF FWHM was $4.7''$ in imaging mode and $\sim 6.7''$ in spectroscopic mode.

\begin{figure}
   \resizebox{1.56in}{!}{\includegraphics{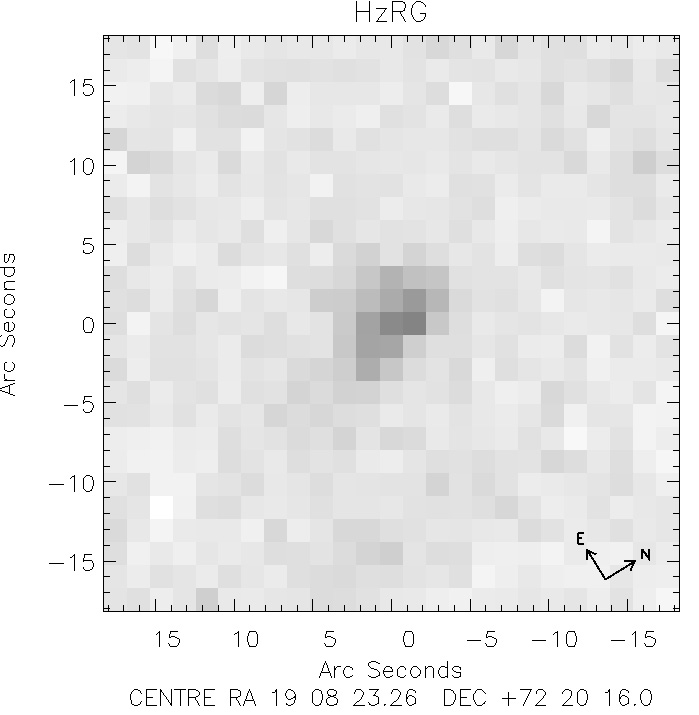}}
   \resizebox{1.54in}{!}{\includegraphics{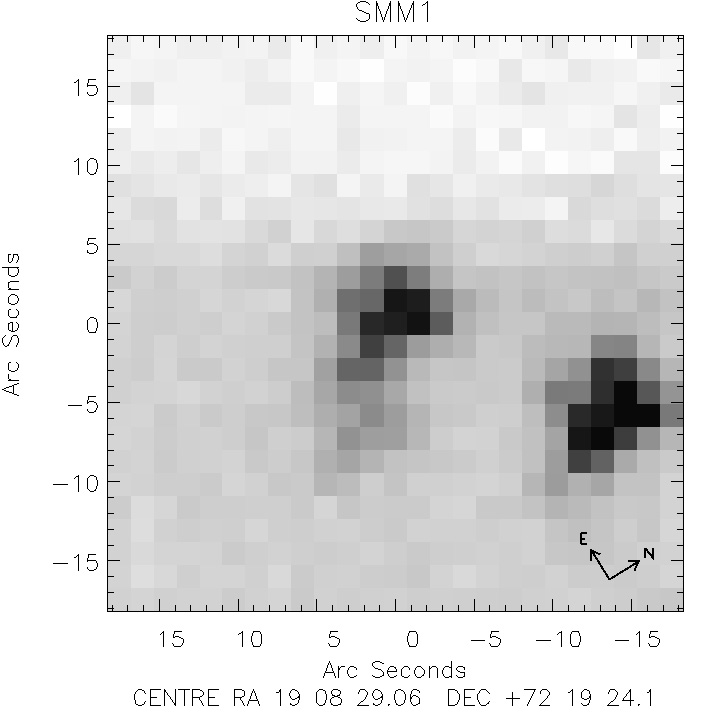}}
          \resizebox{3.0in}{0.02in}{\includegraphics{fig_white.jpg}} % white space
   \resizebox{1.56in}{!}{\includegraphics{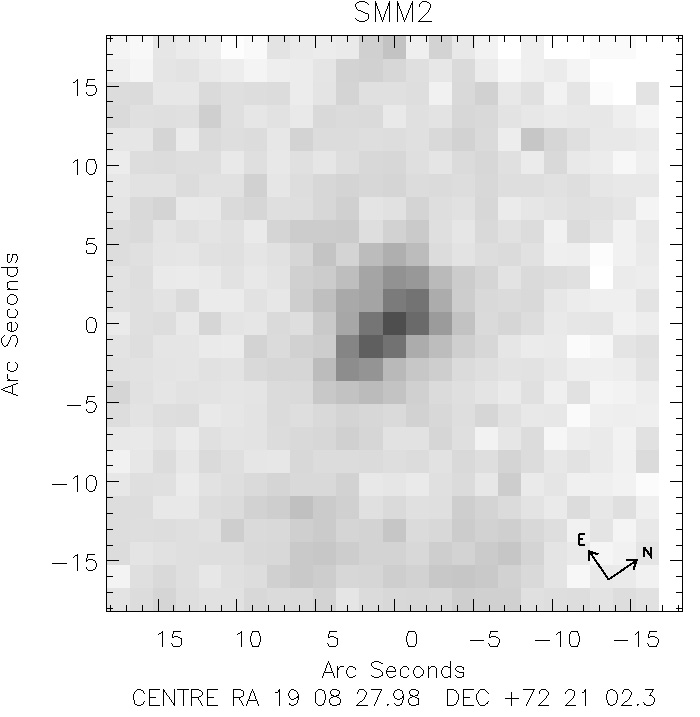}}
       \resizebox{0.08in}{0.5in}{\includegraphics{fig_white.jpg}} % white space
   \resizebox{1.56in}{!}{\includegraphics{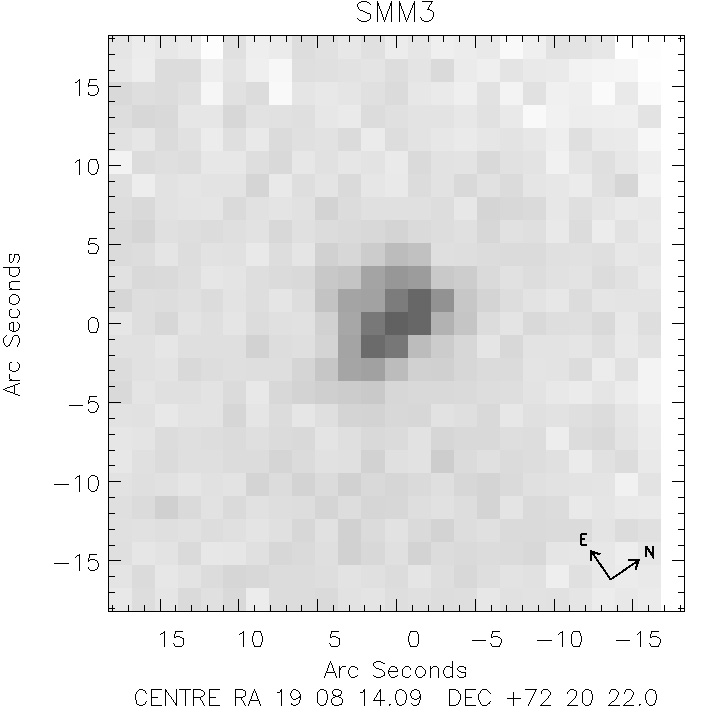}}         
\caption{AKARI 3 $\mu$m images of the 8C1909+722 radio galaxy and its companion submillimetre galaxies. The figures are centred on the positional centroids of the sepctra, and each is a median stack of the pointings taken for each source using the N3 filter. The dispersion direction is horizontal and the FWHM of the kernel used in the source extraction is equivalent to a width of 2 and 3 pixels in the wavelength and spatial directions respectively.}\label{fig:images}
\end{figure}

\begin{figure}
  \begin{center}  
 %    \resizebox{1.72in}{!}{\includegraphics{image_4C6007_rev5a.eps}}  
 \resizebox{1.8in}{!}{\includegraphics{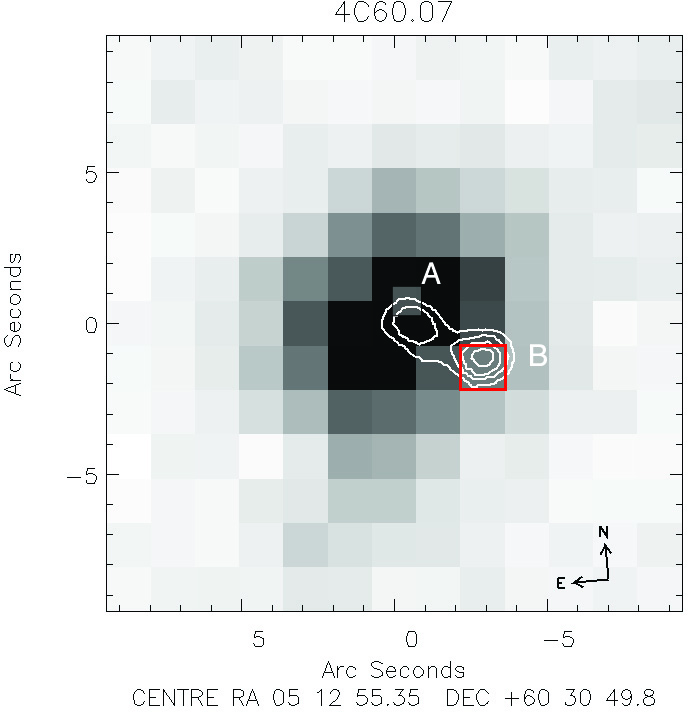}} %% alternative to previous with centre coords
\caption{\small A blow-up of the {\it AKARI} 3 $\mu$m image of 4C60.07, overlaid by contours (at 3.5, 4.5, 5.5 and 6.5 $\sigma$) from the SMA 870 $\mu$m image for the centre of 4C60.07 from Ivison et al.\ (2008) which resolve the two submillimetre sources close to the HzRG. The labelling of the submillimetre sources  \textquoteleft A\textquoteright ~and  \textquoteleft B\textquoteright ~follows Ivison et al.\ (2008). The radio source overlaps \textquoteleft A\textquoteright ~and is centred $<0.5\arcsec$~north-east of it. A red square marks the positional centre of the {\it AKARI} spectrum. The dispersion direction is horizontal and the FWHM of the kernel used in the source extraction was equivalent to a width of 2 and 3 pixels in the wavelength and spatial directions respectively.} \label{fig:images_blowup}
 \end{center} 
\end{figure}

We selected the spectroscopy Astronomical Observing Template (AOT)
{\it IRCZ4} in the configuration {\it b;Np} (which selects the grism and the  point source aperture). In Phase III,  this configuration gives a 5$\sigma$ detection limit for point sources of $\sim 2$ mJy and a line sensitivity of $\sim$ 5$\times$10$^{-18}$ Wm$^{-2}$ for each pointing. Each pointing consisted of a reference image (see Figures \ref{fig:images} and \ref{fig:images_blowup}) and at least 8  exposure frames for the {\it NG} grism, bracketed by 10 dark frames (see Onaka et al.\ 2009 for details). Each frame exposure was $\sim$70 seconds, giving an integrated exposure time for each pointing of $\sim$9.3 minutes. We carried out 10 pointings per source wherever possible. %Given the nature of {\it AKARI's} orbit, target visibility is a strong function of ecliptic latitude constraining possible target selection, especially close to the ecliptic plane.

%Based on extrapolation of the line strengths of H$\alpha$ in local ULIRGs (Veilleux et al.\ 1997; Dannerbauer et al.\ 2005),  

At the redshifts of the targets considered in this paper, the H$\alpha$ hydrogen recombination line falls within our 2.5 $\mu$m - 5.0 $\mu$m observed wavelength range. However, H$\beta$ was outside our  range. No other emission lines were detected. We used the reference image of the larger 10$\arcmin$ $\times$ 10$\arcmin$ N3 band field which is attached to the 1$\arcmin$ $\times$ 1$\arcmin$ grism field, smoothed with a 5 $\times$ 5 median boxcar, to confirm the identification of our sources with images from public archives.

\subsection{Data reduction pipeline}\label{sec:reduction}

The IRC data reduction pipeline for the warm phase (Onaka et al.\ 2009) was originally used to analyse our data. However, we found the correction for spacecraft jitter between the $\sim$8 spectroscopic frames within each pointing and sky subtraction did not yield satisfactory results, so we wrote our own pipeline using Interactive Data Language ({\sc idl}) to reduce the raw spectroscopic data. 

Our new pipeline includes dark subtraction, saturation masking (physical detector saturation is detected by scaled values from a short exposure and masked out), wavelength calibration and spectral response calibration using the calibration data from the IRC pipeline. We wrote our own routines to handle sky subtraction in which we fitted a sixth-order polynomial in the dispersion direction to remove a banding pattern across the frame, and a second-order polynomial in the image direction. We also wrote routines to handle de-glitching and to estimate the offsets between frames for each pointing caused by spacecraft jitter, using the simultaneous 10$\arcmin\times$10$\arcmin$ imaging data. Our pipeline included {\sc idl} routines for zerofootprint drizzling into an grid expanded by 5 times in each direction and for noise-weighted feature extraction which were previously developed for the SCUBA Half Degree Extragalactic Survey (SHADES, Serjeant et al.\ 2008), using a gaussian with a FWHM kernel in the expanded grid of 10 and 15 pixels in the wavelength and spatial directions respectively. This equates to a FWHM kernel of 2 and 3 pixels in the original data. We also wrote an {\sc idl} graphic user interface (GUI) routine to visualize various elements of the reduction on an interactive basis. Finally, the results were stacked by coadding the noise-weighted pointings for each target. The reduced spectra had lower noise and  fewer artefacts than those from the {\it AKARI} pipeline. 

The spectroscopic flat fields provided in the calibration data have low signal-to-noise and were found to increase the rms of our data slightly, so they were not used, following guidance in the manual (Onaka et al.\ 2009).

\begin{table*}
%\tabletypesize{\scriptsize}
%\tablewidth{0pt}
\caption{Estimates of star formation rates and line widths. Estimates of SFR(H$\alpha$) for the SMGs are based on H$\alpha$ emission luminosities, and deconvolved FWHM are based on the width of the H$\alpha$ lines for the HzRGs. No adjustment for extinction has been made. FIR luminosities and SFRs are based on the 850 $\mu$m flux (see Table \ref{table:targets}) assuming an SMMJ2135-0102 SED. SFR(H$\alpha$)s for the HzRGs are not quoted due to line contamination from their AGNs. Redshifts in italics were identified by this work.}\label{sfr}
\begin{center}
   \scalebox{0.95}{
\begin{tabular}{lccccccrc}
\hline

                                         &  Redshift      & Flux(H$\alpha$)              &SNR(H$\alpha$)    & L$_{\rm H\alpha}$   & SFR$_{\rm H\alpha}$            &L$_{\rm FIR}$   & SFR$_{\rm FIR}$  & FWHM(H$\alpha$)  \\
Source                            & (z)                 & (10$^{-19}$Wm$^{-2}$) &                                   &(10$^{36}$W)           & (${\rm M}_{\odot}$yr$^{-1}$)  & (10$^{13}$L$_{\sun}$)     & (${\rm M}_{\odot}$yr$^{-1}$)  & (km s$^{-1}$)\\
\hline
\it{HzRGs}  \\
8C1909+722 HzRG   &3.536             &    ~~7.9                              &  ~3.1                            &  8.6  $\pm$2.8         &    ...                                        & 2.5 $\pm$0.2       &  ~4300 $\pm$1500      &  9400 $\pm 1600$ \\
4C60.07 HzRG            & 3.788           &    ~~2.8                              &  ~2.5                             &   3.6 $\pm$1.4        &   ...                                         & 1.7 $\pm$0.3       & ~2900 $\pm$1000       &   4800 $\pm 1000$ \\
{\it SMGs}\\
8C1909+722 SMM1   & {\it 3.536}    &    ~~3.1                               &  10.0                            &   3.3 $\pm$0.3        & 260 $\pm$~80                      & 1.6 $\pm$0.2        &  ~2800 $\pm$1000       &   ...  \\
8C1909+722 SMM2    & {\it 3.536}   &    ~~3.5                               &  ~2.4                             &  3.8 $\pm$1.6        & 300 $\pm$100                    & 0.6 $\pm$0.2        &  ~1100 $\pm$~500       &   ...   \\
8C1909+722 SMM3    &...                  &   $<$0.02                           & ...                                  &   $<$ 0.02                &  ...                                           & 0.3 $\pm$0.2        & ~~500 $\pm$~300        &  ...   \\

\hline
\end{tabular}
}
\end{center}
\end{table*}

\section{Results}\label{sec:results}

We show in Figure \ref{fig:spectra} the spectra of the five targeted sources, four of which appear to have significant H$\alpha$ emission. The spectra have been stacked by co-adding the noise-weighted pointings for each source. These spectra had been smoothed in the source extraction routine during the data reduction described in \S \ref{sec:reduction},  and were not otherwise smoothed. The signal-to-noise ratios (SNRs) of these detections are 3.1 and 2.5 for the HzRGs and 10.0 and 2.4 for the submillimetre galaxies (see Table \ref{sfr}). Although some of the  SNRs are relatively modest, the redshifts for the two HzRGs were already known from Ly-$\alpha$ and other emission lines (see \S \ref{sec:previous}) and all four  H$\alpha$ line identifications were observed at  the wavelengths corresponding closely to these redshifts. The estimated noise as a function of wavelength was calculated by splitting the raw data for each source into two halves and subtracting the two spectra obtained.

\subsection{Spectra of the HzRGs}

The spectrum for the radio galaxy 8C1909+722 HzRG has a broad H$\alpha$ emission line at the correct wavelength for its redshift, as shown in Figure \ref{fig:spectra} (top). The spectrum was taken from the centre of the target.

For the radio galaxy 4C60.07, Figure \ref{fig:images_blowup} shows the {\it AKARI} 3 $\mu$m image overplotted with the SMA $870\,\mu$m contours from  Ivison et al.\ (2008). There are two submillimetre peaks, the weakest (labelled \textquoteleft A\textquoteright) coincident with the $3\,\mu$m peak, and the strongest (\textquoteleft B\textquoteright) offset by about $3''$. The stacked spectrum for this radio source has a broad H$\alpha$ emission line at \textquoteleft B\textquoteright  ~(see Figure \ref{fig:spectra}, bottom). We did not detect an H$\alpha$ line at \textquoteleft A\textquoteright.

The  FWHM  of the H$\alpha$ lines for the two HzRGs are shown in Table \ref{sfr} and are 9400$\pm{1600}$  km s$^{-1}$ and 4800$\pm{1000}$ km s$^{-1}$ respectively, levels which show that dust-shrouded quasars are present in these sources.

%The  FWHM  of the H$\alpha$ lines for the two HzRGs are shown in Table \ref{sfr} and are 8800$^{+~800}_{-1400}$  km s$^{-1}$ and 4900$^{+500}_{-500}$ km s$^{-1}$ respectively, levels which show that dust-shrouded quasars are present in these sources.

\subsection{Spectra of the submillimetre galaxies near  8C1909+722}

SMM1 shows considerable structure, both in the submillimetre (see Figure \ref{fig:images_submm}) and at 3 $\mu$m (see Figure \ref{fig:images}). The second-brightest of the submillimetre peaks shows an H$\alpha$ emission line, but the other peaks do not. The emission line confirms that SMM1 is at the same redshift as the radio galaxy. 

\begin{figure}
  \begin{center}  

\resizebox{2.92in}{1.0in}{\includegraphics{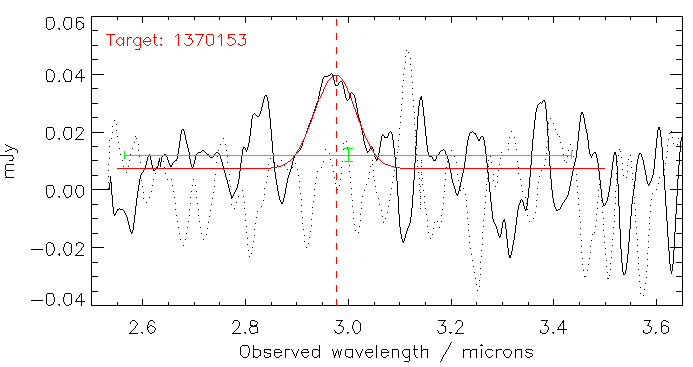}}
     \resizebox{2.5in}{0.03in}{\includegraphics{fig_white.jpg}}   % white space
     
\resizebox{2.91in}{1.03in}{\includegraphics{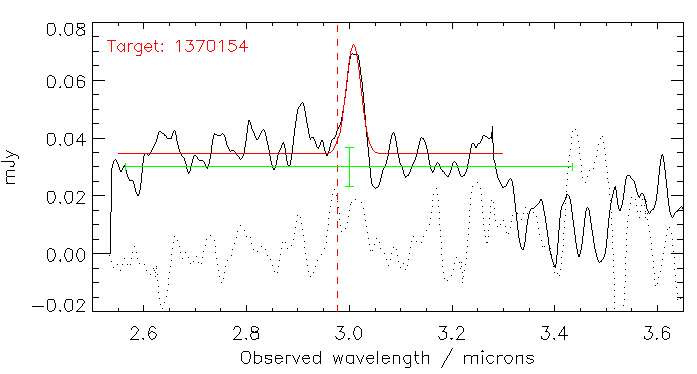}}
     \resizebox{2.5in}{0.03in}{\includegraphics{fig_white.jpg}}
     
\resizebox{2.92in}{1.03in}{\includegraphics{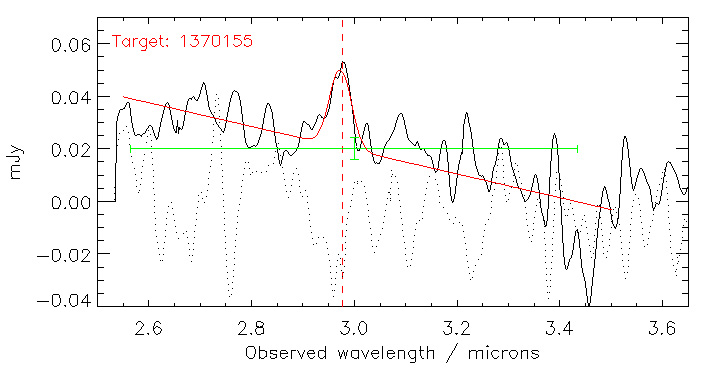}}
     \resizebox{2.5in}{0.03in}{\includegraphics{fig_white.jpg}}
     
\resizebox{2.9in}{1.03in}{\includegraphics{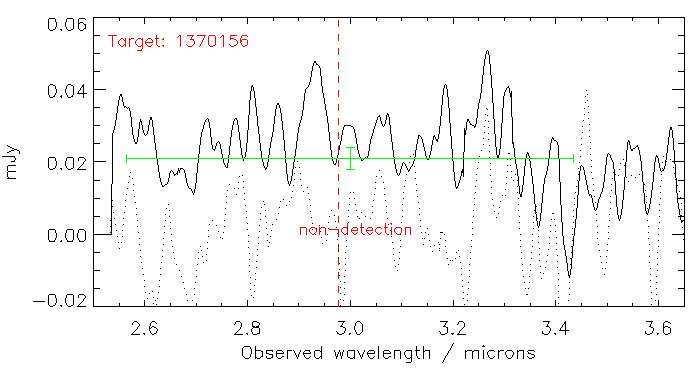}}
    \resizebox{2.5in}{0.03in}{\includegraphics{fig_white.jpg}}
    
\resizebox{2.9in}{1.04in}{\includegraphics{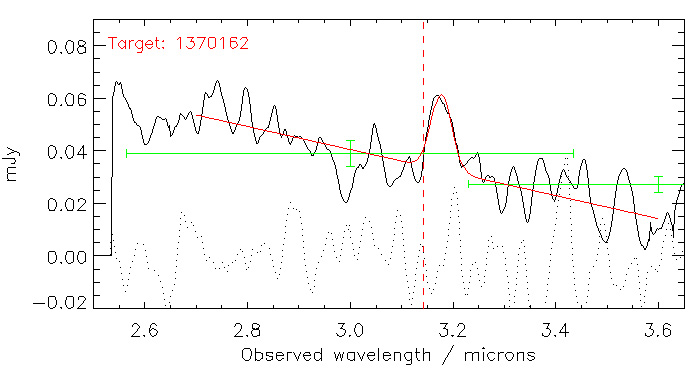}}

\caption{\small From top:  8C1909+722 HzRG, SMM1, SMM2, SMM3 (non-detection) and 4C60.07. The dashed red vertical lines show the position of H$\alpha$ emission lines at the expected redshifts. The solid red lines show the gaussian least-squares fits to the H$\alpha$ line and continuum in the wavelength region shown. The dotted grey lines show the noise levels as a function of wavelength, calculated as described in the text. The green points at 3.0 $\mu$m are from the {\it AKARI} broadband photometry (the reference image). The green point at 3.6 $\mu$m for 4C60.07 is from {\it Spitzer} IRAC broadband photometry. {\it Spitzer} photometry for 8C1909+722 is an order of magnitude higher and may include flux from another source; 0.8 $\mu$m Keck photometry (De Breuck et al.\ 2001) is the same order of magnitude as the {\it AKARI} data. }\label{fig:spectra}
 \end{center} 
\end{figure}

SMM2 also shows H$\alpha$ emission (see Figure \ref{fig:spectra}), again at the same redshift ($z=3.536$), confirming that this galaxy is also associated with the radio galaxy. The spectrum for SMM2 is taken from the centre of the target in the {\it AKARI} 3 $\mu$m image.

SMM3 does not show a peak at the expected wavelength, although there is a strong peak about 0.05 $\mu$m lower. We have not taken this as a convincing H$\alpha$ emission line. Unlike SMM1 and SMM2, this galaxy is not aligned with the radio jets.% and was not detected at 350 $\mu$m (Greve et al.\ 2006).

Stevens et al.\ (2003) suggested that the radio galaxy, the nearby submillimetre galaxies and other clumps had formed as a single galaxy cluster; we have confirmed that this association is correct in the case of two of these submillimetre sources.

 %The red point at 3.0 $\mu$m flux for the three SMMs are from the {\it AKARI} reference image, scaled by the relative {\it Spitzer/AKARI} levels of the HzRGs.

\subsection{Star Formation Rates}\label{sec:sfr}
We have used the H$\alpha$ emission lines to estimate the star formation rates of the submillimetre galaxies, using the formula from Kennicutt (1998):

\begin{equation}
{\rm SFR}/(M_{\odot}{\rm yr}^{-1})=7.9\times10^{-35}{L_{H\alpha} /W}
\end{equation}

\noindent assuming a Salpeter IMF and solar abundances.  The results for the two submillimetre galaxies for which we detected H$\alpha$ lines are $260\pm80$ M$_\odot$ yr$^{-1}$ and $300\pm100$ M$_\odot$ yr$^{-1}$ respectively (see Table \ref{sfr}). No adjustment has been made for dust extinction. 

%\begin{equation}
%{\rm SFR}/({\rm M}_{\odot}{\rm yr}^{-1})=7.9\times10^{-35}{{\rm L_{H\alpha}} /{\rm W}}
%\end{equation}

We can obtain an alternative measure of the SFRs by using the 850 $\mu$m luminosity (Table \ref{table:targets}) to estimate the 60 $\mu$m luminosity, then using the 60 $\mu$m luminosity to obtain an estimate of the far-infrared luminosity for 8-1000~$\mu$m ($L_{\rm FIR}$). We have assumed the SED of the submillimetre galaxy SMMJ2135-0102 (the \textquoteleft Eyelash\textquoteright) for these estimates, and then used the formula for estimating the star formation rate from Kennicutt (1998):

\begin{equation}
{\rm SFR}/(M_{\odot}{\rm yr}^{-1})=  4.5\times10^{-37}{L_{\rm FIR} /W}
\end{equation}

The results of this calculation are shown in Table \ref{sfr}.

%\begin{equation}
%{\rm SFR}/({\rm M}_{\odot}{\rm yr}^{-1})= \epsilon^{-1} \times 4.5\times10^{-37}{{\rm L_{FIR}} /{\rm W}}
%\end{equation}

The SFRs estimated from H$\alpha$ are lower than the SFRs estimated from L$_{\rm{FIR}}$ by factors of about 11 and 4 respectively, suggesting dust obscuration. Swinbank et al.\ (2004) found an SFR(FIR)/SFR(H$\alpha$) ratio of $\sim 10$  in a  study of submillimetre to radio galaxies in the redshift range to $z=1.408$  to $z=2.692$. Takata et al.\ (2006) found  close agreement between the two methods of estimating SFRs after adjusting the H$\alpha$-based estimates for dust extinction by an average factor of 2.9$\pm$0.5 in a study of submillimetre-selected ULIRGs at $0.9<z<2.7$. Figure \ref{fig:LHaz} shows that our SMGs have luminosities and dust extinction at comparable levels with these earlier studies. 

The total FIR luminosity calculated above also shows that SMM1 and the two HzRGs are HyperLuminous InfraRed Galaxies (L$_{\rm{FIR}}>10^{13}$$L_{\sun}$), and SMM2 is a ULIRG.

\section{Discussion and Conclusions}\label{sec:discussion}

These are the first (tentative) detections of H$\alpha$ emission lines in HzRGs / SMGs at $z>2.8$.

We have found that the H$\alpha$ lines in the two HzRGs are broad. Broad H$\alpha$ emission lines in HzRGs are not rare (Larkin et al.\ 2000; Nesvadba et al.\ 2011) and indicate that these sources should be classified as reddened quasars rather than galaxies (see the discussion in Rawlings et al.\ 1995), a situation that can lead to misleading estimates of stellar populations in the host galaxy. This is a particular problem in studies using near-/mid-infrared photometry of sources which may otherwise appear to be narrow-line AGNs. %The confirmation that these radio sources are quasars also provides direct confirmation of the simplest AGN unification models (e.g.\ Antonucci 1993; Urry \& Padovani 1995); indeed this is the first reddened broad-line confirmation of AGN unification in a radio galaxy at $z>3.5$.

For 4C60.07, our detection of a broad H$\alpha$ line was at the location of the submillimetre source \textquoteleft B\textquoteright ~(see Figure \ref{fig:images_blowup}).  Ivison et al.\ (2008) suggested that \textquoteleft B\textquoteright ~was a gas-rich starburst/AGN on the basis of the red mid-infrared colours. Our discovery that \textquoteleft B\textquoteright ~is a quasar supports the argument that binary AGNs at close separations may be due to the triggering or enhancement of AGN activity during mergers. The unexpectedly high prevalence of binary quasars at high redshifts (Djorgovski 1991; Hennawi et al.\ 2010) provided strong support for this idea. A recent study of binary quasars found that simple halo occupation distribution models under-predict quasar clustering at small separations (Kayo \& Oguri 2012). That such a well-studied source had not previously been shown to be a quasar adds to the evidence that there is a higher fraction of binary AGN/quasars at high redshift than previously realised.

Our H$\alpha$ luminosity and FWHM results for the two HzRGs are both consistent with those of lower redshift radio galaxies (e.g.\ Nesvadba et al.\ 2011).

Our detection of H$\alpha$ emission lines from two submillimetre galaxies in the region of 8C1909+722 provides confirmation of their association with the HzRG. The extent of the system is $\sim$700 kpc (based on Stevens et al.\ 2003), suggesting that this may be evolving into a cluster of galaxies or possibly a single galaxy. The complex SMM1 is $\sim80$ kpc in extent and appears to be in the process of merging. In the 8C1909 field we have the first confirmation of multiple U/HyLIRGs in a protocluster region at $z>3$, giving a combined SFR of $\sim 8000~{\rm M}_{\odot}$yr$^{-1}$. 

\begin{figure}
  \begin{center}  
    \resizebox{3.38in}{2.6in}{\includegraphics{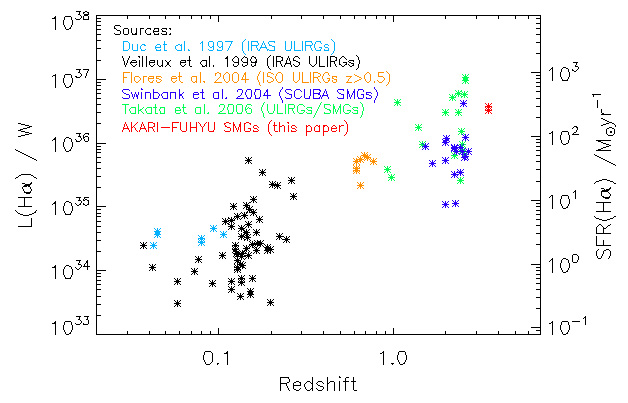}}
      \resizebox{3.15in}{0.22in}{\includegraphics{fig_white.jpg}}   
      \resizebox{3.07in}{!}{\includegraphics{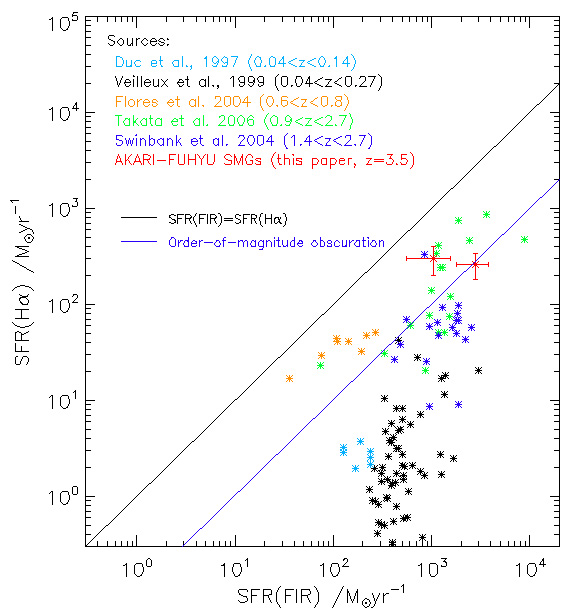}}
\caption{\small Top: H$\alpha$ luminosity for the two SMGs as a function of redshift in comparison with earlier studies. Bottom: Star formation rates based on H$\alpha$ luminosity versus far-infrared luminosity of the two SMGs compared that to those of  recent studies. Takata et al.\ (2006) and Flores et al.\ (2004) data are shown before correction for extinction. Some sources show evidence of AGNs; all are SMGs. Our results are shown in red.}\label{fig:LHaz}
 \end{center} 
\end{figure}

The high levels of star formation in the two SMGs are shown with previous results in Figure \ref{fig:LHaz}. Our H$\alpha$ luminosities and SFR(FIR)/SFR(H$\alpha$) ratios are comparable to those found in recent studies at $0.9<z<2.7$ suggesting similar levels of dust obscuration.

%Eddington bias may be present; also

\vspace{-15 pt}

\section*{Acknowledgments}

This research is based on observations with {\it AKARI}, a JAXA project with participation of ESA. This work was supported by STFC under grant ST/G002533/1. IRS acknowledges support from STFC and the Leverhulme Trust. MI acknowledges support from the Creative Research Initiative program No. 2010-0000712 of NRFK. Thanks to Mark Swinbank for the SMMJ2135-0102 SED. Thanks also to the anonymous referee for many helpful comments.

\clearpage

%% the file, uncomment the \plotone, \plottwo, and
%% If you need a layout that cannot be achieved with \plotone or
%% \plottwo, you can invoke the graphicx package directly with the
%% \includegraphics command or use \plotfiddle. For more information,
%% please see the tutorial on "Using Electronic Art with AASTeX" in the
%% documentation section at the AASTeX Web site, http://aastex.aas.org/
%%
%% The examples below also include sample markup for submission of
%% supplemental electronic materials. As always, be sure to check
%% the instructions to authors for the journal you are submitting to
%% for specific submissions guidelines as they vary from

%% This example uses \plotone to include an EPS file scaled to
%% 80% of its natural size with \epsscale. Its caption
%% has been written to indicate that additional figure parts will be
%% available in the electronic journal.

\label{lastpage}

\end{document}